\documentclass[twoside,12pt]{article}
\usepackage{epsfig}

\newcommand{\be}{\begin{equation}}
\newcommand{\ee}{\end{equation}}
\newcommand{\bea}{\begin{eqnarray}}
\newcommand{\eea}{\end{eqnarray}}

\begin{document}

\title{ \vspace{1cm} Recent developments in effective field theory}
\author{S.\ Scherer\\ \\
Institut f\"ur Kernphysik, Johannes Gutenberg-Universit\"at Mainz,\\
J.~J.~Becher-Weg 45, D-55099 Mainz, Germany}
\date{November 2, 2007}
\maketitle

\begin{abstract}
   We will give a short introduction to the
one-nucleon sector of chiral perturbation theory and will address
the issue of a consistent power counting and renormalization.
   We will discuss the infrared regularization and the
extended on-mass-shell scheme.
   Both allow for the inclusion of
further degrees of freedom beyond pions and nucleons and the
application to higher-loop calculations.
   As applications we consider the chiral expansion of the nucleon mass to order ${\cal
   O}(q^6)$ and the inclusion of vector and axial-vector mesons in the calculation
of nucleon form factors.
\end{abstract}
\section{Introduction}
   Effective field theory (EFT) has become a powerful tool in the description of the
strong interactions at low energies.
   The central idea is due to Weinberg \cite{Weinberg:1978kz}:
   \begin{quote}
"...  if one writes down the most general possible Lagrangian,
including all terms consistent with assumed symmetry principles, and
then calculates matrix elements with this Lagrangian to any given
order of perturbation theory, the result will simply be the most
general possible S--matrix consistent with analyticity, perturbative
unitarity, cluster decomposition and the assumed symmetry
principles."
\end{quote}
    The prerequisite for an effective field theory program
is (a) a knowledge of the most general effective Lagrangian and (b)
an expansion scheme for observables in terms of a consistent power
counting method.
   The application of these ideas to the interactions among the Goldstone bosons
   of spontaneous chiral symmetry breaking in QCD results in
mesonic chiral perturbation theory (ChPT)
\cite{Weinberg:1978kz,Gasser:1983yg} (see, e.g.,
Refs.~\cite{Scherer:2002tk,Scherer:2005ri,Bijnens:2006zp,Bernard:2007zu}
for an introduction and overview).
   In the following we will outline some recent developments in devising
a renormalization scheme leading to a simple and consistent power
counting for the renormalized diagrams of a manifestly
Lorentz-invariant approach to baryon ChPT \cite{Gasser:1987rb}.

\section{Renormalization and Power Counting}
\subsection{Effective Lagrangian and Power Counting}

  The effective Lagrangian relevant to the one-nucleon sector
consists of the sum of the purely mesonic and $\pi N$ Lagrangians,
respectively,
\begin{displaymath}
{\cal L}_{\rm eff}={\cal L}_{\pi}+{\cal L}_{\pi N}={\cal
L}_\pi^{(2)}+{\cal L}_\pi^{(4)}+\cdots+{\cal L}_{\pi N}^{(1)}+{\cal
L}_{\pi N}^{(2)}+\cdots,
\end{displaymath}
which are organized in a derivative and quark-mass expansion
\cite{Weinberg:1978kz,Gasser:1983yg,Gasser:1987rb}.
   For example, the lowest-order basic Lagrangian ${\cal L}_{\pi N}^{(1)}$,
already expressed in terms of renormalized parameters and fields, is
given by
\begin{equation}
\label{LpiN1}
{\cal L}_{\pi N}^{(1)}=\bar \Psi \left( i\gamma_\mu
\partial^\mu - m\right) \Psi
-\frac{1}{2}\frac{\texttt{g}_A}{F} \bar \Psi \gamma_\mu \gamma_5
\tau^a \partial^\mu \pi^a \Psi +\cdots,
\end{equation}
where $m$, $\texttt{g}_A$, and $F$ denote the chiral limit of the
physical nucleon mass, the axial-vector coupling constant, and the
pion-decay constant, respectively.
   The ellipsis refers to terms containing external fields and
higher powers of the pion fields.
   When studying higher orders in perturbation theory one encounters
ultraviolet divergences.
   As a preliminary step, the loop integrals are regularized,
typically by means of dimensional regularization.
   For example, the simplest dimensionally regularized integral
relevant to ChPT is given by
\begin{eqnarray*}
I(M^2,\mu^2,n)&=&\mu^{4-n}\int\frac{\mbox{d}^nk}{(2\pi)^n}\frac{i}{k^2-M^2+i0^+}
=\frac{M^2}{16\pi^2}\left[
R+\ln\left(\frac{M^2}{\mu^2}\right)\right]+O(n-4),
\end{eqnarray*}
where $R=\frac{2}{n-4}-[\mbox{ln}(4\pi)+\Gamma'(1)]-1$ approaches
infinity as $n\to 4$.
   The 't Hooft parameter $\mu$ is responsible for the fact that the integral has
the same dimension for arbitrary $n$.
   In the process of renormalization the
counter terms are adjusted such that they absorb all the ultraviolet
divergences occurring in the calculation of loop diagrams.
   This will be possible, because we include in the Lagrangian all
of the infinite number of interactions allowed by symmetries
\cite{Weinberg:1995mt}.
   At the end the regularization is removed by taking the limit
$n\to 4$.
   Moreover, when renormalizing, we still have the freedom of choosing a renormalization
prescription.
   In this context we will adjust the finite pieces of the renormalized couplings such that
renormalized diagrams satisfy the following power counting:
   a loop integration in $n$ dimensions counts as $q^n$,
pion and fermion propagators count as $q^{-2}$ and $q^{-1}$,
respectively, vertices derived from ${\cal L}_{\pi}^{(2k)}$ and
${\cal L}_{\pi N}^{(k)}$ count as $q^{2k}$ and $q^k$, respectively.
   Here, $q$ collectively stands for a small quantity such as the pion
   mass, small external four-momenta of the pion, and small external
three-momenta of the nucleon.
   The power counting does not uniquely fix the renormalization scheme,
i.~e.~there are different renormalization schemes leading to the
above specified power counting.

\begin{figure}[tb]
\begin{center}
\includegraphics[height=2cm]{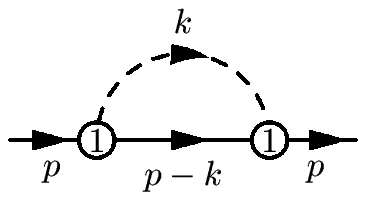}
\caption{One-loop contribution to the nucleon self energy. The
number 1 in the interaction blobs refers to ${\cal L}_{\pi
N}^{(1)}$.} \label{fig:nucleonselfenergypionloop}
\end{center}
\end{figure}

   As an example, consider the one-loop
contribution of Fig.\ \ref{fig:nucleonselfenergypionloop} to the
nucleon self energy.
   After renormalization, we would like
to have the order $D=n\cdot 1-2\cdot 1-1+1\cdot 2=n-1.$
   The application of the $\widetilde{\rm MS}$ renormalization scheme of ChPT
\cite{Gasser:1983yg,Gasser:1987rb}---indicated by ``r''---yields
\begin{displaymath}
\Sigma_{\rm loop}^r=-\frac{3 \texttt{g}_{A}^2}{4 F^2}\left[
-\frac{M^2}{16\pi^2}(p\hspace{-.4em}/\hspace{.1em}+m)
+\cdots\right]= {\cal O}(q^2),
\end{displaymath}
where $M^2=2B\hat m$ is the lowest-order expression for the squared
pion mass in terms of the low-energy coupling constant $B$ and the
average light-quark mass $\hat m$ \cite{Gasser:1983yg}.
   The $\widetilde{\rm MS}$-renormalized result does not
produce the desired low-energy behavior which, for a long time, was
interpreted as the absence of a systematic power counting in the
relativistic formulation of ChPT.

\subsection{Infrared Regularization and Extended On-Mass-Shell Scheme}
   Several methods have been suggested to obtain a consistent power
counting in a manifestly Lorentz-invariant approach.
   We will illustrate the underlying ideas in terms of a typical one-loop integral
   in the chiral limit
\begin{displaymath}
H(p^2,m^2;n)= \int \frac{{\mbox d}^n k}{(2\pi)^n}
\frac{i}{[(k-p)^2-m^2+i0^+][k^2+i0^+]},
\end{displaymath}
where $\Delta=(p^2-m^2)/m^2={\cal O}(q)$ is a small quantity.
   Applying the dimensional counting analysis of
Ref.~\cite{Gegelia:1994zz} the result of the integration is of the
form
\begin{displaymath}
H\sim F(n,\Delta)+\Delta^{n-3}G(n,\Delta),
\end{displaymath}
where $F$ and $G$ are hypergeometric functions which are analytic
for $|\Delta|<1$ for any $n$.

   In the infrared regularization of Becher and Leutwyler \cite{Becher:1999he}
one makes use of the Feynman parametrization
\begin{displaymath}
{1\over ab}=\int_0^1 {\mbox{d}z\over [az+b(1-z)]^2}
\end{displaymath}
with $a=(k-p)^2-m^2+i0^+$ and $b=k^2+i0^+$.
   The resulting integral over the Feynman parameter $z$ is then rewritten as
\begin{eqnarray*}
H=\int_0^1 \mbox{d}z \cdots &=& \int_0^\infty \mbox{d}z \cdots
- \int_1^\infty \mbox{d}z \cdots,\\
\end{eqnarray*}
where the first, so-called infrared (singular) integral satisfies
the power counting, while the remainder violates power counting but
turns out to be regular and can thus be absorbed in counter terms.

   The central idea of the extended on-mass-shell (EOMS)
scheme \cite{Fuchs:2003qc} consists of subtracting those terms which
violate the power counting as $n\to 4$.
   Since the terms violating the power counting are analytic in small
quantities, they can be absorbed by counter term contributions.
   In the present case, we want the renormalized integral to be of
the order $D=n-1-2=n-3$.
   To that end one first expands the integrand in
small quantities and subtracts those integrated terms whose order is
smaller than suggested by the power counting.
   The corresponding subtraction term reads
\begin{displaymath}
H^{\rm subtr}=\int \frac{\mbox{d}^n k}{(2\pi)^n}\left.
\frac{i}{[k^2-2p\cdot k +i0^+][k^2+i0^+]}\right|_{p^2=m^2}
\end{displaymath}
and the renormalized integral is written as $ H^R=H-H^{\rm
subtr}={\cal O}(q) $ as $n\to 4$.

\subsection{Remarks}
\begin{itemize}
\item Using a suitable renormalization condition one
obtains a consistent power counting in manifestly Lorentz-invariant
baryon ChPT including, e.g., vector mesons \cite{Fuchs:2003sh} or
the $\Delta(1232)$ resonance \cite{Hacker:2005fh} as explicit
degrees of freedom.
\item  The infrared regularization of Becher and Leutwyler
\cite{Becher:1999he} has been reformulated in a form analogous to
the EOMS renormalization \cite{Schindler:2003xv}.
\item The application of both
infrared and extended on-mass-shell renormalization schemes to
multi-loop diagrams was explicitly demonstrated by means of a
two-loop self-energy diagram \cite{Schindler:2003je}.
   In both cases the renormalized diagrams satisfy a
straightforward power counting.
\end{itemize}

\section{Applications}

\subsection{Chiral Expansion of the Nucleon Mass to Order ${\cal O}(q^6)$}

Using the reformulated infrared regularization
\cite{Schindler:2003xv} we have calculated the nucleon mass up to
and including order ${\cal O}(q^6)$ in the chiral expansion
\cite{Schindler:2006ha,Schindler:2007dr}:
\begin{eqnarray}\label{H1:emff:MassExp}
    m_N &=& m +k_1 M^2 +k_2 \,M^3 +k_3 M^4 \ln\frac{M}{\mu}
+ k_4 M^4  + k_5 M^5\ln\frac{M}{\mu} + k_6 M^5  \nonumber\\&& + k_7
M^6 \ln^2\frac{M}{\mu}+ k_8 M^6 \ln\frac{M}{\mu} + k_9 M^6.
\end{eqnarray}
   In Eq.~(\ref{H1:emff:MassExp}), $m$ denotes the nucleon mass in the
chiral limit, $M^2$ is the leading term in the chiral expansion of
the square of the pion mass, $\mu$ is the renormalization scale; all
the coefficients $k_i$ have been determined in terms of infrared
renormalized parameters.
   The results of Ref.\ \cite{Schindler:2006ha} represent the first complete
two-loop calculation in manifestly Lorentz-invariant baryon
ChPT.
   Our results for the renormalization-scheme-independent terms agree
with the heavy-baryon ChPT results of Ref.~\cite{McGovern:1998tm}.

\begin{figure}[tb]
\begin{center}
\includegraphics[width=0.45\textwidth, angle=0]{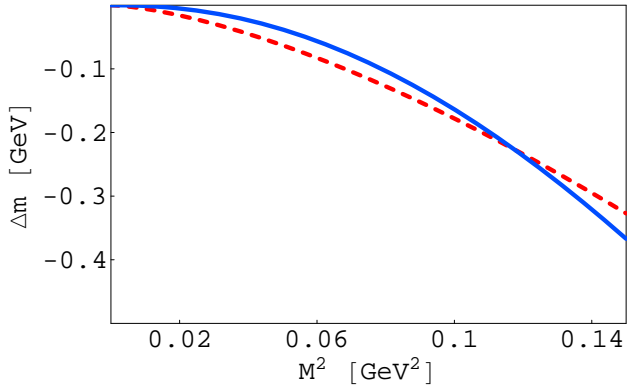}
\caption{\label{fig:nucleonmass} Pion mass dependence of the term
$k_5 M^5 \ln(M/m_N)$ (solid line) for $M<400$ MeV. For comparison
also the term $k_2 M^3$ (dashed line) is shown.}
\end{center}
\end{figure}

   The numerical contributions from higher-order terms cannot be
calculated so far since, starting with $k_4$, most expressions in
Eq.~(\ref{H1:emff:MassExp}) contain unknown low-energy coupling
constants (LECs) from the Lagrangians of order ${\cal O}(q^4)$ and
higher.
   The coefficient $k_5$ is free of higher-order LECs
and is given in terms of the axial-vector coupling constant
$\texttt{g}_A$ and the pion-decay constant $F$.
   While the values for both $\texttt{g}_A$ and $F$ should be taken in the chiral
limit, we evaluate $k_5$ using the physical values $g_A=1.2695(29)$
and $F_\pi=92.42(26)$ MeV.
   Setting $\mu=m_N$, $m_N=(m_p+m_n)/2=938.92$ MeV, and $M=M_{\pi^+}=139.57$ MeV
we obtain $k_5 M^5 \ln(M/m_N) = -4.8$ MeV.
   This amounts to approximately $31$\% of the leading nonanalytic contribution
at one-loop order, $k_2 M^3$.
   Figure \ref{fig:nucleonmass} shows the pion mass dependence of the term
$k_5 M^5 \ln(M/m_N)$ (solid line) in comparison with the term $k_2
M^3$ (dashed line) for $M<400$ MeV.
   For $M\approx 360\,\mbox{MeV}$ the $k_5$ term is as large as the
   $k_2$ term.

\subsection{Electromagnetic Form Factors}
   Imposing the relevant symmetries such as translational invariance,
Lorentz covariance, the discrete symmetries, and current
conservation, the nucleon matrix element of the electromagnetic
current operator $J^\mu(x)$ can be parameterized in terms of two
form factors,
\begin{equation}
\label{H1:emff:empar} \langle N(p')|J^{\mu}(0)|N(p)\rangle=
\bar{u}(p')\left[F_1^N(Q^2)\gamma^\mu
+i\frac{\sigma^{\mu\nu}q_\nu}{2m_p}F_2^N(Q^2) \right]u(p),\quad
N=p,n,
\end{equation}
   where $q=p'-p$, $Q^2=-q^2$, and $m_p$ is the proton mass.
   At $Q^2=0$, the so-called Dirac and Pauli form factors $F_1$ and
$F_2$ reduce to the charge and anomalous magnetic moment in units of
the elementary charge and the nuclear magneton $e/(2m_p)$,
respectively,
\begin{displaymath}
F_1^{p}(0)=1,\quad F_1^{n}(0)=0,\quad F_2^{p}(0)=1.793,\quad
F_2^{n}(0)=-1.913.
\end{displaymath}
   The Sachs form factors $G_E$ and $G_M$ are linear combinations of $F_1$ and
$F_2$,
\begin{displaymath}
G_E^N(Q^2)=F_1^N(Q^2)-\frac{Q^2}{4m_p^2}F_2^N(Q^2),\quad
G_M^N(Q^2)=F_1^N(Q^2)+F_2^N(Q^2), \quad N=p,n,
\end{displaymath}
and, in the non-relativistic limit, their Fourier transforms are
commonly interpreted as the distribution of charge and magnetization
inside the nucleon.
   The description of the electromagnetic form factors of the nucleon
presents a stringent test for any theory or model of the strong
interactions.

\begin{figure}[tb]
\begin{center}
\epsfig{file=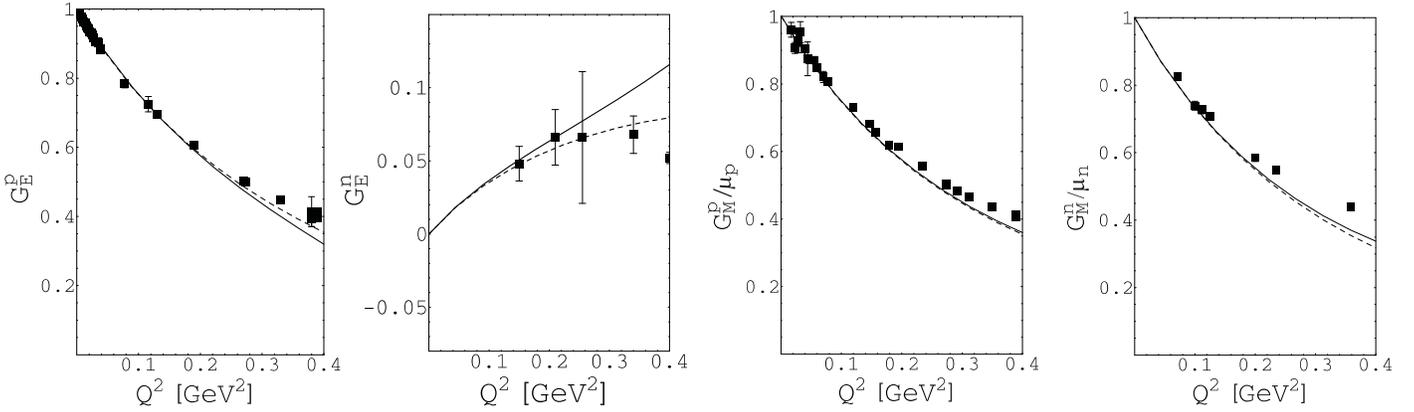,width=\textwidth}
\caption{\label{H1:emff:G:neu} The Sachs form factors of the nucleon
in manifestly Lorentz-invariant chiral perturbation theory at ${\cal
O}(q^4)$ including vector mesons as explicit degrees of freedom.
Full lines: results in the extended on-mass-shell scheme; dashed
lines: results in infrared regularization. The experimental data are
taken from Ref.\ \cite{Friedrich:2003iz}.}
\end{center}
\end{figure}

   Calculations in Lorentz-invariant baryon ChPT up to
fourth order fail to describe the proton and nucleon form factors
for momentum transfers beyond $Q^2\sim 0.1\, \mbox{GeV}^2$
\cite{Kubis:2000zd,Fuchs:2003ir}.
   In Ref.\ \cite{Kubis:2000zd} it was shown that the
inclusion of vector mesons can result in the re-summation of
important higher-order contributions.
   In Ref.\ \cite{Schindler:2005ke} the electromagnetic form factors of the
nucleon up to fourth order have been calculated in manifestly
Lorentz-invariant ChPT with vector mesons as explicit degrees of
freedom.
   A systematic power counting for the renormalized diagrams has been
implemented using both the extended on-mass-shell renormalization
scheme and the reformulated version of infrared regularization.
    The inclusion of vector mesons results in a
considerably improved description of the form factors (see
Fig.~\ref{H1:emff:G:neu}).
    The most dominant contributions come from
tree-level diagrams, while loop corrections with internal vector
meson lines are small \cite{Schindler:2005ke}.

\subsection{Axial and Induced Pseudoscalar Form Factors}
   Assuming isospin symmetry, the most general
parametrization of the isovector axial-vector current evaluated
between one-nucleon states is given by
\begin{equation}\label{H1_axff_FFDef}
\langle N(p')| A^{\mu,a}(0) |N(p) \rangle = \bar{u}(p')
\left[\gamma^\mu\gamma_5 G_A(Q^2) +\frac{q^\mu}{2m_N}\gamma_5
G_P(Q^2) \right] \frac{\tau^a}{2}u(p),
\end{equation}
where $q=p'-p$, $Q^2=-q^2$, and $m_N$ denotes the nucleon mass.
   $G_A(Q^2)$ is called the axial form factor and
$G_P(Q^2)$ is the induced pseudoscalar form factor.
     The value of the axial form factor at zero momentum transfer is defined as
the axial-vector coupling constant, $g_A=G_A(Q^2=0) =1.2695(29)$,
and is quite precisely determined from neutron beta decay.
   The $Q^2$ dependence of the axial form factor can be obtained
either through neutrino scattering or pion electroproduction.
   The second method makes use of the so-called Adler-Gilman relation
\cite{Adler:1966gd} which provides a chiral Ward identity
establishing a connection between charged pion electroproduction at
threshold and the isovector axial-vector current evaluated between
single-nucleon states (see, e.g., Ref.\
\cite{Bernard:2001rs,Fuchs:2003vw} for more details).
   The induced pseudoscalar form factor $G_P(Q^2)$ has been investigated in
ordinary and radiative muon capture as well as pion
electroproduction (see Ref.\ \cite{Gorringe:2002xx} for a review).

\begin{figure}[tb]
\begin{minipage}[b]{0.3\textwidth}
\includegraphics[width=\textwidth]{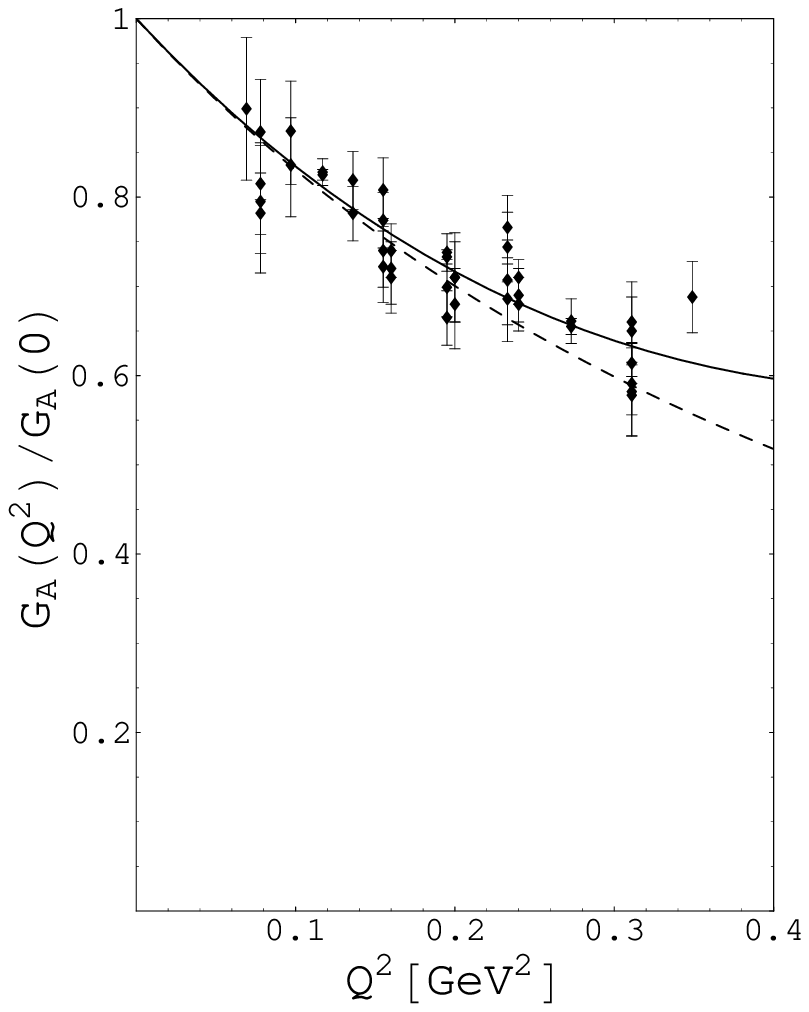}
\end{minipage}
\hspace{2em}
\begin{minipage}[b]{0.5\textwidth}
\includegraphics[width=\textwidth]{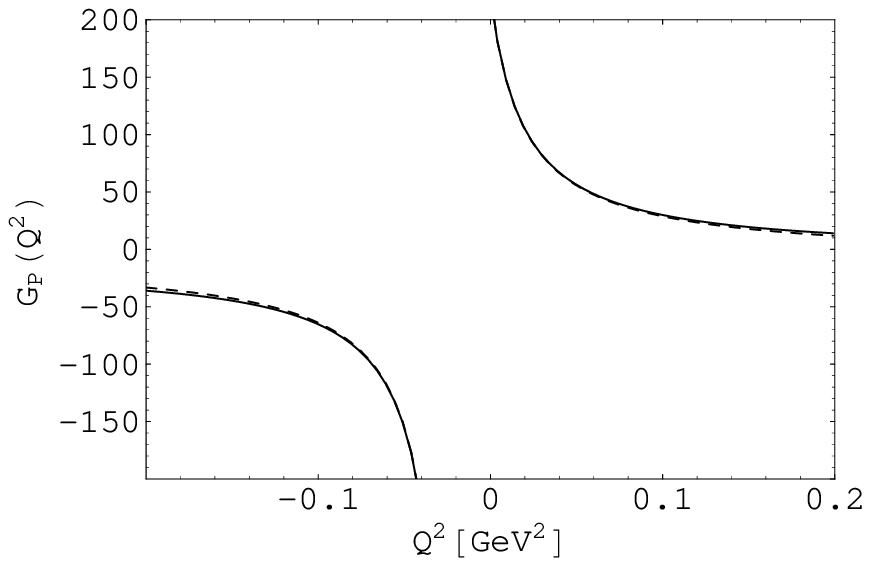}
\end{minipage}
\caption{\label{H1_axff_GAwith}Left panel: Axial form factor $G_A$
in manifestly Lorentz-invariant ChPT at ${\cal O}(q^4)$ including
the axial-vector meson $a_1(1260)$ explicitly. Full line: result in
infrared renormalization, dashed line: dipole parametrization. The
experimental data are taken from Ref.\ \cite{Bernard:2001rs}.
 Right
panel:
   The induced pseudoscalar form factor $G_P$ in manifestly Lorentz-invariant ChPT
at ${\cal O}(q^4)$ including the axial-vector meson $a_1(1260)$
explicitly. Full line: result with axial-vector meson; dashed line:
result without axial-vector meson.
   One can clearly see the dominant pion pole contribution at $Q^2\approx
   -M_\pi^2$.}
\end{figure}

   In Ref.\ \cite{Schindler:2006it} the form factors $G_A$ and
$G_P$ have been calculated in manifestly Lorentz-invariant baryon
ChPT up to and including order ${\cal O}(q^4)$.
   In addition to the standard treatment including the
nucleon and pions, the axial-vector meson $a_1(1260)$ has also been
considered as an explicit degree of freedom.
   The inclusion of the axial-vector meson effectively results in one additional
low-energy coupling constant which has been determined by a fit to
the data for $G_A(Q^2)$.
   The inclusion of the axial-vector meson results in an improved
description of the experimental data for $G_A$ (see
Fig.~\ref{H1_axff_GAwith}), while the contribution to $G_P$ is
small.

\subsection{Pion-Nucleon Form Factor}
   The pion-nucleon form factor $G_{\pi N}(Q^2)$ may be defined in terms of
the pseudoscalar quark density $P^a=i\bar{q}\gamma_5 \tau^a q$ and
the average light-quark mass $\hat m$ as \cite{Gasser:1987rb}
\begin{equation}
\label{GpiN} \hat m \langle N(p')| P^a (0) | N(p) \rangle =
         \frac{M_\pi^2 F_\pi}{M_\pi^2 + Q^2}
         G_{\pi N}(Q^2)i\bar{u}(p') \gamma_5 \tau^a u(p),
\end{equation}
where $q=p'-p$, $Q^2=-q^2$, and $\Phi^a(x)\equiv \frac{\hat m P^a
(x)}{M_\pi^2 F_\pi}$ is the corresponding interpolating pion field.
   The pion-nucleon coupling constant is given by
$g_{\pi N}=G_{\pi N}(-M_\pi^2)$.
   Using the (QCD-) partially conserved axial-vector current (PCAC) relation,
$\partial_\mu A^{\mu,a}= \hat m P^a$, the pion-nucleon form factor
is completely given in terms of the axial and the induced
pseudoscalar form factors,
\begin{displaymath}
   2m_N G_A(Q^2) - \frac{Q^2}{2m_N} G_P(Q^2) =
       2\frac{M_\pi^2 F_\pi}{M_\pi^2 + Q^2} G_{\pi N}(Q^2).
\end{displaymath}
   This is an exact relation which holds true for any value of
   $Q^2$.
   The result at ${\cal O}(q^4)$ is given by \cite{Schindler:2006it}
\begin{displaymath}
\label{GpiNresult} G_{\pi N}(Q^2)=\frac{m_N g_A}{F_\pi}-g_{\pi
N}\Delta \frac{Q^2}{M_\pi^2} + \cdots
\end{displaymath}
where $\Delta=1-\frac{m_N g_A}{F_\pi g_{\pi N}}$ denotes the
Goldberger-Treiman discrepancy.
   The chiral expansion of the pion-nucleon coupling constant can
   be found in Ref.\ \cite{Schindler:2006it}.

\begin{figure}[tb]
\begin{center}
\includegraphics[width=0.5\textwidth, angle=0]{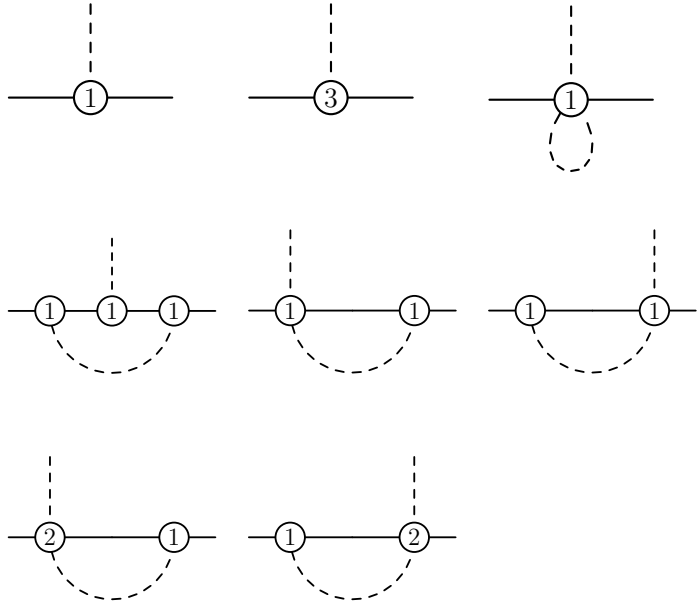}
\caption{\label{fig:piNNvertex} Diagrams contributing to the
pion-nucleon vertex at ${\cal O}(q^4)$.}
\end{center}
\end{figure}

   Finally, we would like to stress that one carefully has to distinguish
between the pion-nucleon form factor $G_{\pi N}(Q^2)$ of Eq.\
(\ref{GpiN}) on the one hand and the renormalized pion-nucleon
vertex in ChPT on the other hand.
   The unrenormalized vertex, when evaluated between on-shell nucleon states, is
   of the form
\begin{equation}
\Gamma_5(q^2)\gamma_5 \tau^a.
\end{equation}
   At ${\cal O}(q^4)$ one needs to calculate the diagrams shown in Fig.\
   \ref{fig:piNNvertex}.
   In general, the pion-nucleon vertex depends on the choice of the field
variables in the effective Lagrangian.
   Only at $q^2=M_\pi^2$, we
expect the same coupling strength, since both $\hat m
P^a(x)/(M_\pi^2 F_\pi)$ and the field $\pi^a$ of Eq.\ (\ref{LpiN1})
serve as interpolating pion fields:
\begin{displaymath}
G_{\pi N}(-M_\pi^2)=g_{\pi N}=Z_\Psi\sqrt{Z_\pi}\Gamma_5(M_\pi^2).
\end{displaymath}
   On the other hand, for arbitrary $q^2$
\begin{displaymath}
 G_{\pi N}(Q^2)\neq Z_\Psi\sqrt{Z_\pi}\Gamma_5(q^2).
\end{displaymath}
   In the present case, the pion-nucleon vertex is only an auxiliary
quantity, whereas the ``fundamental'' quantity (entering chiral Ward
identities) is the matrix element of the pseudoscalar density.

\section{Summary and Conclusion}

   Both the infrared regularization and the
EOMS scheme allow for a simple and consistent power counting in
manifestly Lorentz-invariant baryon chiral perturbation theory.
   We have discussed some results of a two-loop calculation of
   the nucleon mass.
   The inclusion of vector and axial-vector mesons as explicit
degrees of freedom leads to an improved phenomenological description
of the electromagnetic and axial form factors, respectively.
   Work on the application to electromagnetic processes such as Compton
scattering and pion production is in progress.

   I would like to thank D.~Djukanovic, T.~Fuchs, J.~Gegelia,
 G.~Japaridze, and M.~R.~Schindler for the
fruitful collaboration on the topics of this talk. This work was
made possible by the financial support from the Deutsche
Forschungsgemeinschaft (SFB 443 and SCHE 459/2-1) and the EU
Integrated Infrastructure Initiative Hadron Physics Project
(contract number RII3-CT-2004-506078).

\end{document}